\documentclass[11pt]{article}
\usepackage[T1]{fontenc}
\usepackage{lmodern}
\usepackage{microtype}
\usepackage[margin=1in]{geometry}
\usepackage{amsmath,amssymb,bm}
\usepackage{graphicx}
\usepackage{booktabs,longtable,array}
\usepackage{cite}
\usepackage{url}
\usepackage[hidelinks]{hyperref}
\usepackage{doi}
\usepackage{xcolor}
\hypersetup{
  pdftitle={SymPerturb converts symptom-network structure into testable intervention priorities},
  pdfauthor={Zheng Zhu, Junwen Yu, Tiantian Hu, Zhongfang Yang, Jiaqing Wang}
}

\title{SymPerturb converts symptom-network structure into testable intervention priorities}
\author{Zheng Zhu\textsuperscript{1,4,*}, Junwen Yu\textsuperscript{2,4}, Tiantian Hu\textsuperscript{1,4}, Zhongfang Yang\textsuperscript{3,4}, Jiaqing Wang\textsuperscript{4}}
\date{}

\begin{document}
\maketitle
\begin{center}
\small
\textsuperscript{1}School of Nursing, Fudan University, Shanghai, China\\
\textsuperscript{2}NYU Shanghai, Shanghai, China\\
\textsuperscript{3}School of Nursing, Soochow University, Suzhou, China\\
\textsuperscript{4}Yulin AI-Enhanced HealthCare Lab, Shanghai, China\\[0.4em]
\textsuperscript{*}Correspondence: \href{mailto:zhengzhu@fudan.edu.cn}{zhengzhu@fudan.edu.cn}
\end{center}

\begin{abstract}
Symptom networks encode conditional dependence but do not by themselves identify causal or clinically actionable intervention targets. We introduce SymPerturb, a virtual-perturbation framework that distinguishes four primitive perturbation operators---virtual knockout, virtual knockdown, edge-level communication blocking and node-centred communication blocking---from three analytic procedures---virtual dosage perturbation, combination perturbation and sequence optimisation. The reference Gaussian implementation is embedded in a general location--scale map with symptom-specific target anchors, making explicit that zero anchoring and linked mean--variance attenuation are modelling choices. Seven utility outcomes quantify downstream efficacy, dose efficiency, breadth, cross-module reach, communication blocking, combination value and responsiveness; robustness is reported separately as an uncertainty diagnostic. Their direction-aligned, within-candidate-set weighted mean defines the virtual perturbation priority score (VPPS), which is a relative ranking rather than a transportable clinical utility score. In a known 22-node, four-module generating network, analytical efficacy agreed with 100,000-draw Monte Carlo estimates within 0.0024 standard deviations. The reported finite-sample VPPS results were generated with the original eight-component exploratory score and therefore require regeneration under the revised seven-utility-dimension definition. These simulations provide internal computational verification under model compatibility, not causal or external validation. SymPerturb is intended to generate auditable target hypotheses for longitudinal and experimental testing.
\end{abstract}

\section{Introduction}\label{introduction}

Network models have changed the way multivariate symptom data are
represented. Instead of treating symptoms as interchangeable indicators
of one latent disease, a symptom network represents them as interacting
components whose conditional dependence may be summarised by a Gaussian
graphical model (GGM) \cite{borsboom2013,borsboom2021,epskamp2018,williams2020}. This representation has encouraged the
use of strength, expected influence, bridge centrality and
predictability to identify symptoms that appear structurally important
\cite{robinaugh2016,jones2021,haslbeck2018}. Yet structural importance is not the same quantity as
intervention value. A symptom may be highly connected because it is
downstream of several common causes, may have strong but harmful and
beneficial links that cancel, or may occupy a central position only
within a particular choice of network boundary \cite{hallquist2021,neal2023,neal2022}.

The gap becomes explicit when the scientific question changes from
description to a model-based perturbation query. Descriptive centrality
asks where a node sits in a fitted graph. A perturbation query asks how
a prespecified fitted distribution or graph functional changes under an
explicit operator. It does not ask what would happen under a real
treatment unless the intervention semantics, temporal ordering and
causal assumptions are independently justified. The query therefore
requires three objects that centrality does not supply: a perturbation
operator, a rule for updating the fitted system and a clinically
interpretable outcome functional. Simulation-based intervention analysis
and network-control approaches have begun to provide these objects
\cite{lunansky2022,henry2022,blanken2019}, but their interpretation remains conditional on the
generating model. In particular, conditioning on or statistically
shifting a symptom in a cross-sectional undirected network is not, by
itself, a causal do-intervention \cite{pearl2009}.

Here we formalise SymPerturb, a symptom-network virtual-perturbation
framework, and retain the seven component names used in the original
implementation. These components are not all objects of the same
mathematical type. Virtual knockout, virtual knockdown, edge-level
communication blocking and node-centred communication blocking are
primitive perturbation operators. Virtual dosage perturbation is an
intensity--response evaluation, combination perturbation is a
joint-target construction and sequence optimisation is a decision
procedure over ordered target sets. We also make three assumptions
explicit. First, the zero endpoint is meaningful only after symptom
orientation and clinical anchoring. Second, the reference state operator
links mean attenuation to scale attenuation, which is a modelling choice
rather than a necessary feature of symptom improvement. Third, exact
knockout should be evaluated on a prespecified estimand rather than by
re-estimating a full covariance matrix that retains a constant column.

SymPerturb is intended as a methods contribution. We derive the
reference state update from a Gaussian regression decomposition, express
topology blocking as matrix operators, define seven utility outcomes and
one uncertainty diagnostic, and evaluate finite-sample recovery using
data generated from a known network. The simulation is designed for
internal computational verification: whether analytical and sampled
responses agree, whether population rankings are recovered under model
compatibility, how rankings change across analysis settings and when
exact knockout breaks a downstream estimator. It does not establish
external validity, clinical efficacy or causal transportability. Any
ranking remains a model-derived hypothesis requiring longitudinal,
experimental and patient-centred validation.

\section{The SymPerturb architecture}\label{the-symperturb-architecture}

SymPerturb contains a state layer, a topology layer and a multi-target
layer (Fig. 1). The state layer contains two primitive operators:
virtual knockout (vKO), which anchors a target to a prespecified state
and may remove its residual variance, and virtual knockdown (vKD), which
produces a partial location--scale change. Virtual dosage perturbation
(vDP) evaluates an intensity--response curve generated by a state
operator. The topology layer contains edge-level communication blocking
and node-centred communication blocking, which attenuate one edge or all
edges incident to a target, respectively. The multi-target layer
contains combination perturbation and sequence optimisation; the latter
searches ordered target sets under an explicit decision objective rather
than inferring biological time from a cross-sectional network.

Each primitive operator returns either a distributional change or a
topological change. The three analytic procedures evaluate, combine or
optimise those operators. Seven utility outcomes then reduce the
resulting changes to target-level quantities, while robustness is
retained as a separate uncertainty diagnostic. VPPS is calculated only
after the utility outcomes have been direction-aligned and normalised
across the prespecified candidate set. This taxonomy prevents a category
error: an edge weight, a model-implied symptom shift, a stable rank and
an optimisation policy are different quantities. It also avoids
mechanistic overstatement: ``communication blocking'' is retained as the
framework label but denotes a graph attenuation operation, not evidence
that symptoms literally exchange a biological signal.

\begin{figure}[p]
\centering
\includegraphics[width=\textwidth]{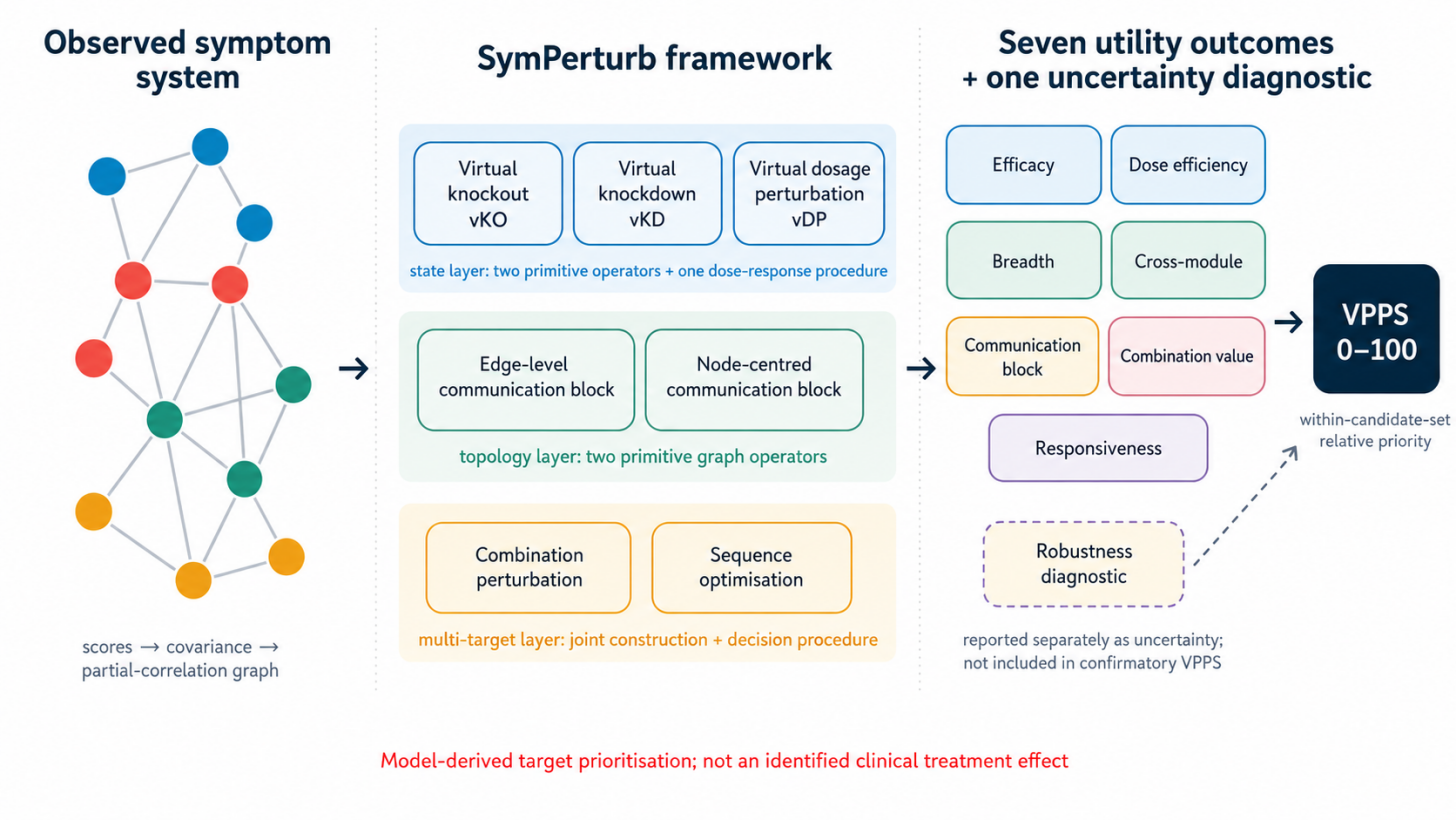}
\caption{The SymPerturb workflow. Virtual perturbation specifies what is changed, how the fitted system is updated and how the response is evaluated. The original component names are retained, but their mathematical roles are distinguished: vKO and vKD are state operators; vDP is an intensity--response evaluation; edge-level and node-centred communication blocking are topology operators; combination perturbation is a joint-target construction; and sequence optimisation is a decision procedure. VPPS ranks model-derived target hypotheses within the analysed candidate set; robustness is reported separately, and neither quantity is an identified clinical treatment effect.}
\label{fig:workflow}
\end{figure}

\section{A general location--scale state intervention with knockout as an endpoint}\label{a-general-locationscale-state-intervention-with-knockout-as-an-endpoint}

Let the symptom vector be partitioned into a target set S and the
remaining symptoms K. Under a multivariate Gaussian model, the
non-target vector has the regression representation

\begin{equation}\label{eq:1}
\mathbf{X}_{K} = \mathbf{\mu}_{K} + \mathbf{B}\left( \mathbf{X}_{S} - \mathbf{\mu}_{S} \right) + \mathbf{\varepsilon},\qquad\mathbf{B} = \mathbf{\Sigma}_{KS}\mathbf{\Sigma}_{SS}^{-1},
\end{equation}

where

\begin{equation}\label{eq:2}
\mathbf{\varepsilon} \sim \mathcal{N}\left( \mathbf{0},\mathbf{\Omega} \right),\qquad\mathbf{\Omega} = \mathbf{\Sigma}_{KK} - \mathbf{\Sigma}_{KS}\mathbf{\Sigma}_{SS}^{-1}\mathbf{\Sigma}_{SK}.
\end{equation}

The matrix $\boldsymbol{\Omega}$ is the Schur complement of $\boldsymbol{\Sigma}_{SS}$. A general state
intervention may use symptom-specific target anchors and separate mean
and scale maps, as formalised in Methods. The numerical results reported
here use the reference special case in which all scales are oriented so
that larger values indicate worse symptoms, the target anchor is zero,
and the target mean displacement and standard deviation are both
multiplied by $\mathbf{D}_{\alpha}=\operatorname{diag}(1-\alpha_s)$. Under that linked special case, replace
the target distribution by

\begin{equation}\label{eq:3}
\mathbf{X}_{S}^{\left( \mathbf{\alpha} \right)} = \mathbf{D}_{\alpha}\mathbf{\mu}_{S} + \mathbf{D}_{\alpha}\left( \mathbf{X}_{S} - \mathbf{\mu}_{S} \right).
\end{equation}

The intervention is exactly the baseline model when every $\alpha_s=0$. In the
linked reference case, the target mean displacement from zero and the
target standard deviation both shrink with dose, and the operator
becomes an exact zero-anchored knockout when every $\alpha_s=1$. Substitution
into the Gaussian regression representation gives

\begin{equation}\label{eq:4}
\mathbb{E}\left[ \mathbf{X}_{K}^{\left( \mathbf{\alpha} \right)} \right] = \mathbf{\mu}_{K} + \mathbf{B}\left( \mathbf{D}_{\alpha}\mathbf{\mu}_{S} - \mathbf{\mu}_{S} \right),
\end{equation}

and

\begin{equation}\label{eq:5}
\operatorname{Var}\left( \mathbf{X}_{K}^{\left( \mathbf{\alpha} \right)} \right) = \mathbf{\Omega} + \mathbf{B}\mathbf{D}_{\alpha}\mathbf{\Sigma}_{SS}\mathbf{D}_{\alpha}\mathbf{B}^{\mathsf{T}}.
\end{equation}

This linked location--scale formulation corrects a discontinuity in a
simple hard-conditioning implementation: if a partial dose fixes a
target to a constant, its variance collapses at any non-zero dose.
However, linking mean attenuation and variance attenuation is a strong
modelling assumption rather than a necessary property of treatment.
SymPerturb therefore treats the linked map as a reference implementation
and requires sensitivity analyses using location-only, scale-only and
independently parameterised location--scale maps. It remains a
statistical intervention defined within the fitted Gaussian model, not a
causal intervention identified from cross-sectional data.

\section{Seven components of SymPerturb}\label{seven-components-of-symperturb}

The seven components are summarised in Table 1 and derived in Methods.
Four are primitive perturbation operators, whereas vDP, combination
perturbation and sequence optimisation are an evaluation, a joint-target
construction and a decision procedure, respectively. Virtual knockout is
stated separately because its numerical consequences differ from those
of partial knockdown. For a single target j, the unit-dose reference
operator has

\begin{equation}\label{eq:6}
\mathbb{E}\left[ X_{j}^{\mathrm{vKO}} \right] = 0,\qquad \operatorname{Var}\left( X_{j}^{\mathrm{vKO}} \right) = 0.
\end{equation}

The non-target mean shift is

\begin{equation}\label{eq:7}
\mathbb{E}\left[ \mathbf{X}_{-j}^{\mathrm{vKO}} \right] - \mathbf{\mu}_{-j} = - \mathbf{\Sigma}_{-j,j}\sigma_{jj}^{-1}\mu_{j}.
\end{equation}

The direct implication is computational, not merely conceptual. A
post-knockout data matrix containing a fixed target has a zero-variance
column. Standardising that column is undefined and the full covariance
matrix is singular. Network comparisons after exact vKO must therefore
remove the target, treat its incident edges as absent, or use a
pre-specified soft vKO with a small non-zero variance. These
alternatives answer different questions and should not be interchanged
without sensitivity analysis. In applied data, the knockout anchor must
also be clinically justified; it need not equal zero.

\begin{longtable}{@{}p{0.18\textwidth}p{0.30\textwidth}p{0.44\textwidth}@{}}
\caption{Seven components of SymPerturb. Virtual knockout, virtual knockdown, edge-level communication blocking and node-centred communication blocking are primitive operators. Virtual dosage perturbation, combination perturbation and sequence optimisation are an intensity--response evaluation, a joint-target construction and a decision procedure, respectively.}\label{tab:components}\\
\toprule
\textbf{Component} & \textbf{Reference mathematical action} & \textbf{Mathematical role and primary question} \\
\midrule
\endfirsthead
\toprule
\textbf{Component} & \textbf{Reference mathematical action} & \textbf{Mathematical role and primary question} \\
\midrule
\endhead
\bottomrule
\endfoot
Virtual knockout, vKO & Exact reference case: $\alpha_j=1$, target anchored at $c_j$; $c_j=0$ and zero residual variance in the reported simulation & Primitive state operator: what is the model-implied response to exact target anchoring? \\
Virtual knockdown, vKD & $0<\alpha_j<1$; partially attenuate target location and scale under a prespecified map & Primitive state operator: what response follows a feasible partial target change? \\
Virtual dosage perturbation, vDP & Evaluate $G_j(\alpha)$ over $\alpha\in[0,1]$ for a selected state operator & Intensity--response procedure: is the response linear, thresholded or saturating? \\
Edge-level communication block & $w'_{uv}=(1-q)w_{uv}$ & Primitive topology operator: how does a selected graph functional change when one edge is attenuated? \\
Node-centred communication block & $w'_{jk}=(1-q)w_{jk}$ for every $k\ne j$ & Primitive topology operator: how dependent is the selected graph functional on edges incident to one node? \\
Combination perturbation & Apply a joint operator to $\{j,k\}$ and compare with single-target operators on a common outcome set & Joint-target construction: does a pair add signed value beyond the better single target? \\
Sequence optimisation & Maximise discounted marginal gains over ordered target sets under explicit costs and constraints & Decision procedure: which feasible order maximises the prespecified objective? \\
\end{longtable}

\section{Seven utility outcomes, a robustness diagnostic and VPPS}\label{seven-utility-outcomes-a-robustness-diagnostic-and-vpps}

For target $j$ at dose $\alpha$, let $\Delta_{i\leftarrow j}^{(\alpha)}$ be the standardised
improvement in non-target symptom i, and define the weighted system
benefit

\begin{equation}\label{eq:8}
G_{j}(\alpha) = \frac{\sum_{i \neq j}w_{i}\Delta_{i \leftarrow j}^{(\alpha)}}{\sum_{i \neq j}w_{i}}.
\end{equation}

The default analysis used equal positive symptom weights. Higher values
were oriented to indicate greater potential value for each utility
outcome. Table 2 defines seven utility outcomes and one uncertainty
diagnostic, with denominators, thresholds and common outcome sets made
explicit. Robustness is not a clinical utility dimension and is
therefore excluded from the confirmatory VPPS.

\begin{longtable}{@{}p{0.18\textwidth}p{0.43\textwidth}p{0.31\textwidth}@{}}
\caption{Seven utility outcomes and one robustness diagnostic in SymPerturb. $q$ is the number of modules; $c(j)$ is the module of target $j$; $Q_T$ is a finite-step topology functional; and $\mathcal{T}_j$ is the pre-specified partner set. Robustness is reported separately and is not included in the confirmatory VPPS. The combination score is a model-based incremental value measure, not an additive causal interaction.}\label{tab:outcomes}\\
\toprule
\textbf{Outcome} & \textbf{Definition} & \textbf{Interpretation} \\
\midrule
\endfirsthead
\toprule
\textbf{Outcome} & \textbf{Definition} & \textbf{Interpretation} \\
\midrule
\endhead
\bottomrule
\endfoot
Downstream efficacy score & Full-dose system benefit, $G_j(1)$ & Mean standardised non-target improvement after full vKO; direct target benefit is reported separately \\
Dose-efficiency score & Mean of $G_j(\alpha_k)/\alpha_k$ over the pre-specified doses & Benefit per unit intervention over partial doses \\
Breadth score & Share of non-target improvements satisfying $\Delta_{i\leftarrow j}^{(1)}\ge\tau$ & Proportion of non-target symptoms exceeding an improvement threshold \\
Cross-module score & Share of other modules with mean improvement at least $\tau_m$ & Proportion of other symptom modules reached \\
Robustness diagnostic & One minus the normalised standard deviation of scenario ranks & Sensitivity of target rank across scenarios; uncertainty diagnostic, not included in confirmatory VPPS \\
Communication-block score & Relative loss in $Q_T$ after node-centred blocking & Relative loss of a prespecified finite-step topology functional after node-centred blocking \\
Combination-value score & Mean positive incremental pair value over partner set $\mathcal{T}_j$ in the current exploratory implementation & Incremental value over the better single target; confirmatory analyses should retain signed values \\
Responsiveness score & $\{G_j(\epsilon)-G_j(0)\}/\epsilon$ & Local low-dose sensitivity of the system \\
\end{longtable}

Each raw utility outcome $R_{jm}$ is normalised across the prespecified
candidate targets:

\begin{equation}\label{eq:9}
\mathrm{score}_{jm} = 100\,\frac{R_{jm} - \min_{\ell}R_{\ell m}}{\max_{\ell}R_{\ell m} - \min_{\ell}R_{\ell m}}.
\end{equation}

If a utility dimension is constant, it contributes a neutral score of 50
rather than an undefined value. The confirmatory VPPS is the weighted
mean of the seven utility dimensions:

\begin{equation}\label{eq:10}
\mathrm{VPPS}_{j} = \frac{\sum_{m = 1}^{7}\omega_{m}\mathrm{score}_{jm}}{\sum_{m = 1}^{7}\omega_{m}},\qquad\omega_{m} \geq 0.
\end{equation}

We used equal weights, $\omega_m=1$, for the reference methodological
analysis. These weights are not clinical utilities. A clinical
application should pre-specify weights using patient priorities,
intervention feasibility, safety and decision-analytic considerations,
then report the unaggregated seven-dimensional utility profile and the
separate robustness diagnostic beside VPPS. Because min--max
normalisation is performed within the candidate set, VPPS is a relative
within-analysis ranking and cannot be directly compared across networks,
cohorts or candidate sets. The numerical VPPS results currently reported
in this draft were generated with the original eight-component
exploratory composite and must be regenerated under the revised
seven-dimension confirmatory rule.

\section{Internal computational verification and finite-sample recovery}\label{internal-computational-verification-and-finite-sample-recovery}

We generated a 22-node GGM with four symptom modules, positive and
negative edges, heterogeneous means and heterogeneous variances. The
precision matrix was positive definite by construction and the observed
0--4 scores were obtained by boundary clipping. Population outcomes were
calculated from the known generating parameters. At each of n=250, 500
and 1,000, 200 independent datasets were generated, fitted and ranked.
The independently generated dataset, not the node or Monte Carlo draw,
was the statistical unit. This design provides internal verification
under model compatibility; it does not test model misspecification or
external validity.

Analytical vKO efficacy closely matched direct simulation. Across 22
targets and 100,000 conditional draws per target, the maximum absolute
discrepancy was 0.0024 standard deviations (Fig. 2a). This verifies the
numerical implementation of the linked reference map and
boundary-expectation calculations. It does not verify the clinical truth
of the GGM, the target anchor or the linked mean--scale assumption.

The original exploratory eight-component VPPS ranking improved
consistently with sample size (Fig. 2b and Table 3). Median Spearman
correlation with the population exploratory VPPS rank was 0.70 at n=250,
0.82 at n=500 and 0.88 at n=1,000. The median proportion of true
top-five targets recovered was 0.60, 0.60 and 0.80, respectively (Fig.
2c). Efficacy, dose efficiency and responsiveness had the strongest rank
recovery. Communication blocking and threshold-based breadth and
cross-module scores required larger samples. The robustness diagnostic
was the clear exception: median rank recovery was 0.07, 0.13 and 0.11
across the three sample sizes (Fig. 2d). These numerical results are
retained as an audit trail but must be regenerated after robustness is
removed from the confirmatory VPPS.

\begin{figure}[p]
\centering
\includegraphics[width=0.95\textwidth]{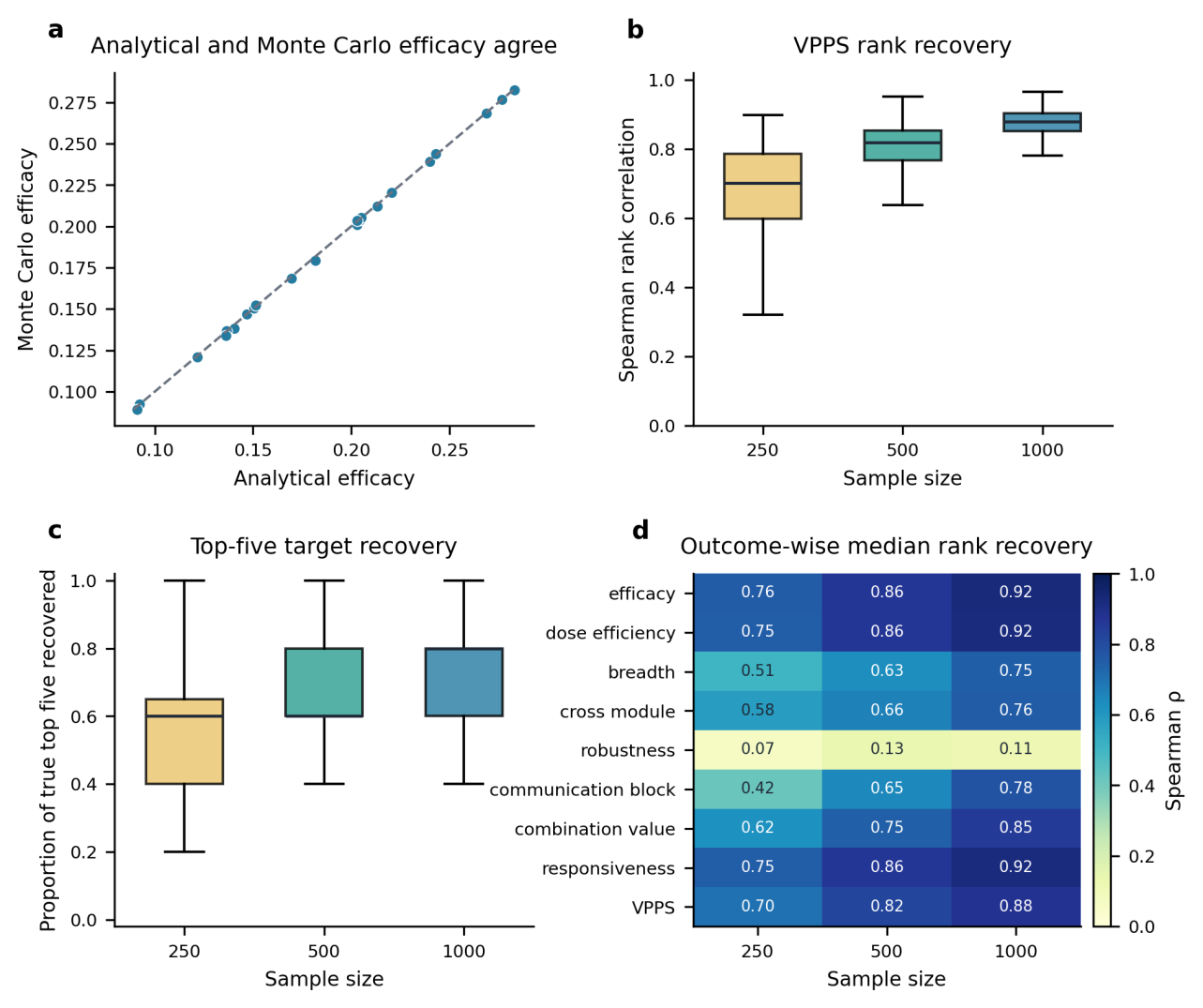}
\caption{Recovery of perturbation outcomes in model-compatible simulated networks. (a) Analytical efficacy versus Monte Carlo efficacy for 22 targets, with the identity line. (b) Distribution of the original exploratory eight-component VPPS rank correlation across 200 independently generated datasets at each sample size. (c) Proportion of the population top five recovered. (d) Median outcome-specific rank correlation. These panels do not contain results for the revised seven-utility-dimension confirmatory VPPS, which requires reanalysis.}
\label{fig:recovery}
\end{figure}

\begin{longtable}{@{}p{0.12\textwidth}p{0.27\textwidth}p{0.23\textwidth}p{0.28\textwidth}@{}}
\caption{Finite-sample recovery of the original exploratory eight-component VPPS. Intervals are the 2.5th and 97.5th percentiles across 200 independently generated datasets, not confidence intervals for a patient population. The table must be regenerated for the revised confirmatory VPPS.}\label{tab:recovery}\\
\toprule
\textbf{Sample size} & \textbf{Median exploratory VPPS8 $\rho$ (95\% Monte Carlo interval)} & \textbf{Median top-five recovery (95\% interval)} & \textbf{Median exploratory VPPS8 RMSE on 0--100 scale (95\% interval)} \\
\midrule
\endfirsthead
\toprule
\textbf{Sample size} & \textbf{Median exploratory VPPS8 $\rho$ (95\% Monte Carlo interval)} & \textbf{Median top-five recovery (95\% interval)} & \textbf{Median exploratory VPPS8 RMSE on 0--100 scale (95\% interval)} \\
\midrule
\endhead
\bottomrule
\endfoot
250 & 0.70 (0.39--0.88) & 0.60 (0.20--0.80) & 16.35 (11.17--23.20) \\
500 & 0.82 (0.65--0.92) & 0.60 (0.40--1.00) & 13.18 (9.47--17.40) \\
1,000 & 0.88 (0.80--0.94) & 0.80 (0.40--1.00) & 10.58 (7.86--13.97) \\
\end{longtable}

\section{Knockout validity, sensitivity and distinction from benchmark rankings}\label{knockout-validity-sensitivity-and-distinction-from-benchmark-rankings}

Exact vKO exposed a deterministic failure mode. The target column had
zero variance in every one of 100 diagnostic replicates; consequently,
full-network standardisation and covariance re-estimation were invalid
in all replicates. Deleting the target and estimating the induced
subgraph was valid in all 100 replicates, as was a soft vKO retaining
5\% of the target standard deviation (Fig. 3b). An exact knockout result
should therefore report the target anchor and whether it represents
state anchoring, graph deletion or a soft intervention.

The population exploratory ranking was insensitive to the 13
pre-specified analysis scenarios. Their Spearman correlations with the
reference provisional ranking had a median of 0.998 and ranged from
0.983 to 1.000 (Fig. 3a). This high population stability did not make
the estimated robustness diagnostic reliable, because small sample
changes could still rearrange near-tied scenario ranks. Population
sensitivity and sampling uncertainty are distinct and should be reported
separately through scenario profiles, bootstrap rank distributions and
top-k selection probabilities.

The 0--4 boundary produced visible curvature in a high-burden saturation
stress test, whereas the corresponding unbounded Gaussian response was
linear (Fig. 3c). Thus efficacy, dose efficiency and responsiveness can
be almost redundant under a linear Gaussian model but diverge under
range restriction or other non-linear measurement functions. Their
empirical correlation should be inspected before they are combined, and
sensitivity analyses should compare location-only, scale-only and linked
state maps.

Finally, the population exploratory VPPS rank was only moderately
associated with absolute strength-centrality rank, \ensuremath{\rho}=0.52 (Fig. 3d). In
this synthetic example, Rumination, Fatigue, Concentration, Functional
limitation and Low energy had the highest population values. These names
are labels in generated data, not clinical recommendations. The contrast
shows that SymPerturb differs from node strength, but comparison with
strength alone is insufficient to establish incremental value. A
confirmatory benchmark should include expected influence, bridge
metrics, predictability, control-based metrics and a burden-only rule
under pre-specified ranking, top-k and decision-regret losses.

\begin{figure}[p]
\centering
\includegraphics[width=0.95\textwidth]{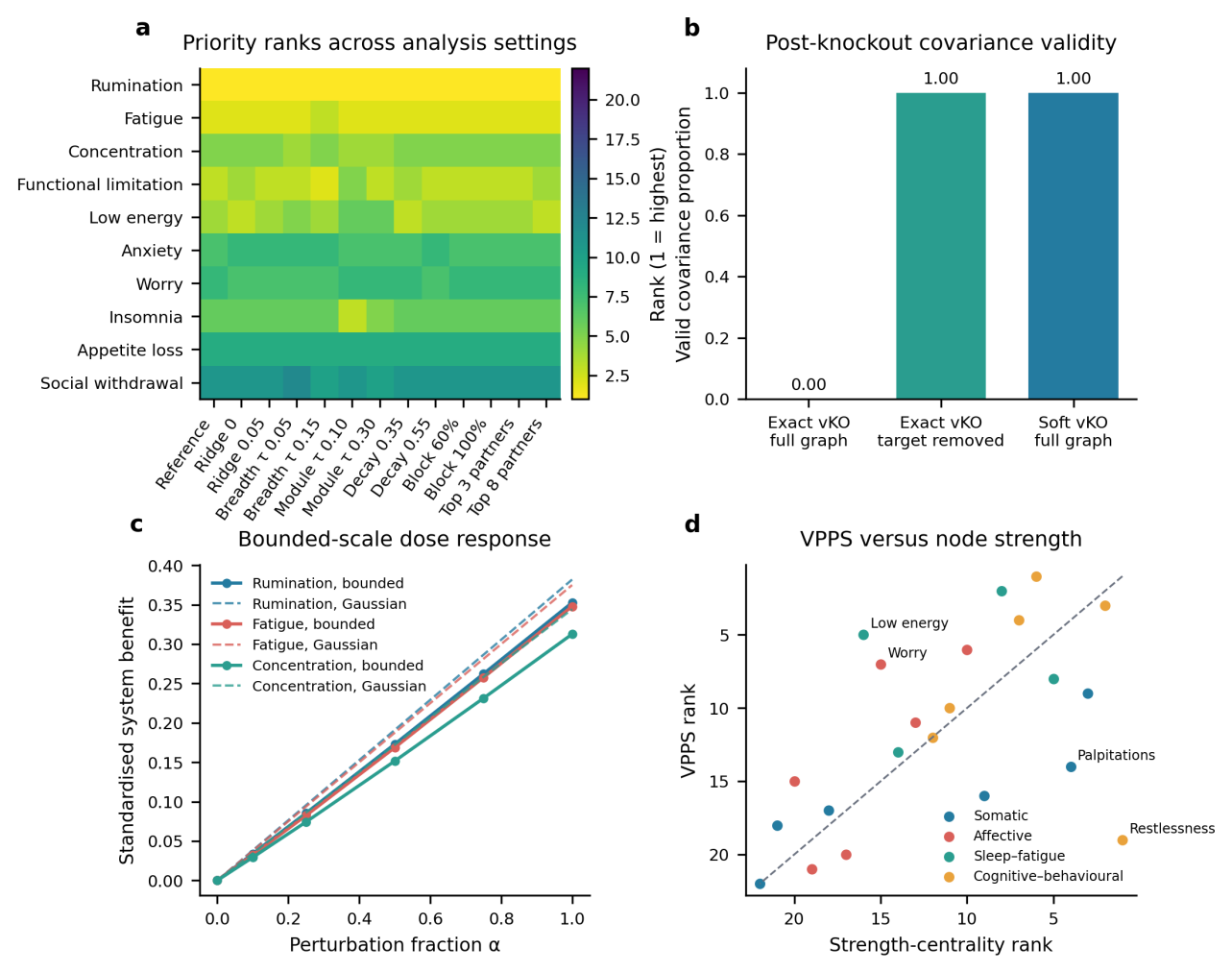}
\caption{Robustness, boundary effects and the centrality contrast. (a) Population target ranks under 13 analysis scenarios. (b) Valid covariance proportion across 100 exact, reduced-network and soft-knockout replicates. (c) Bounded and unbounded dose curves in a high-burden saturation stress test. (d) Population exploratory VPPS rank versus absolute strength-centrality rank; colours denote symptom modules. The strength comparison is illustrative rather than a complete benchmark.}
\label{fig:diagnostics}
\end{figure}

\section{Discussion}\label{discussion}

SymPerturb converts a descriptive symptom network into a set of
explicit, auditable target hypotheses. Its main contribution is not
another centrality statistic. It separates primitive perturbation
operators from the procedures that evaluate, combine or optimise them,
and separates clinical-utility dimensions from estimation uncertainty.
Virtual knockout and knockdown act on symptom distributions;
communication blocking acts on graph topology; virtual dosage
perturbation evaluates an intensity--response curve; combination
perturbation constructs joint targets; and sequence optimisation
searches an ordered decision objective. Seven utility outcomes answer
questions about magnitude, efficiency, reach, topology dependence,
partner value and local sensitivity, while robustness characterises
uncertainty rather than value.

The mathematical formulation clarifies several quantities that can
otherwise be overinterpreted. First, exact vKO is not simply ``100\%
knockdown'' for all downstream computations. It is a degenerate endpoint
at a prespecified anchor and invalidates a full-network covariance
estimate when the target variance is zero. Second, the reference map
assumes that mean displacement and standard deviation contract together;
location-only, scale-only and independently parameterised maps may
represent different intervention semantics. Third, dose efficiency and
responsiveness do not automatically add independent information. In an
unbounded linear Gaussian model, $G_j(\alpha)=\alpha G_j(1)$, so efficacy, per-dose
benefit and low-dose slope are algebraically equivalent. Fourth, the
communication-block score is a property of the chosen topology
functional, including edge thresholding, sign handling, scaling, decay
and path length. It is not direct evidence that symptoms exchange a
biological signal.

The validation provides both support and restraint. Analytical and Monte
Carlo state responses agreed to within 0.0024 standard deviations, and
recovery of the original exploratory VPPS improved with sample size.
Nevertheless, at n=250, the 95\% Monte Carlo interval for rank recovery
extended from 0.39 to 0.88, and top-five recovery could be as low as
0.20. More importantly, the rank-standard-deviation robustness
diagnostic was poorly recovered even at n=1,000. Robustness should
therefore be reported through scenario-specific ranks, bootstrap
selection probabilities and uncertainty intervals rather than included
in VPPS. All composite results must be regenerated using the revised
seven-utility-dimension definition.

Several limitations define the scope of the claims. The current
verification uses one family of synthetic GGMs and an approximately
continuous 0--4 symptom scale. It does not cover strongly ordinal or
zero-inflated data, latent confounding, measurement error, missingness,
longitudinal or person-specific systems, estimator misspecification, or
high-dimensional p/n regimes. The thresholds for breadth and module
reach are analysis choices, not minimal clinically important
differences. Equal VPPS weights are a methodological convention, not a
clinical preference model. The combination score measures improvement
beyond the better single target and, in its current positive-part form,
suppresses harmful joint effects; a confirmatory implementation should
retain signed incremental values. Sequence optimisation uses a
discounted set objective and a restricted candidate search; it does not
solve a general dynamic treatment regime. Most importantly, a
cross-sectional undirected GGM does not identify intervention direction
or remove unmeasured confounding \cite{pearl2009,hallquist2021,neal2023,neal2022}.

A credible translational pathway should therefore proceed in stages. The
first stage is expanded computational validation, including factorial
model-misspecification simulations, complete-pipeline bootstrap
uncertainty, parameter sensitivity, systematic benchmark comparisons and
independent software replication. The second is longitudinal validation,
asking whether natural or treatment-induced changes in a candidate
target precede the predicted downstream changes. The third is
experimental validation with measured target engagement, ideally
followed by prospective decision studies that incorporate safety, cost,
equity and patient preferences. Used within these boundaries, SymPerturb
can narrow a target set and make assumptions testable. It cannot replace
a trial.

\section{Methods}\label{methods}

\subsection{Baseline Gaussian graphical model}\label{baseline-gaussian-graphical-model}

Let $\mathbf{X}\in\mathbb{R}^{p}$ denote continuous or approximately continuous symptom
scores:

\begin{equation}\label{eq:11}
\mathbf{X} \sim \mathcal{N}_{p}\left( \mathbf{\mu},\mathbf{\Sigma} \right),\qquad\mathbf{\Theta} = \mathbf{\Sigma}^{-1}.
\end{equation}

The partial correlation between symptoms i and j, conditional on the
other symptoms, is

\begin{equation}\label{eq:12}
\rho_{ij \cdot - ij} = - \frac{\theta_{ij}}{\sqrt{\theta_{ii}\theta_{jj}}}.
\end{equation}

For finite-sample estimation, the empirical covariance S was diagonally
regularised:

\begin{equation}\label{eq:13}
{\widehat{\mathbf{\Sigma}}}_{\lambda} = \mathbf{S} + \lambda \operatorname{diag}\left( \mathbf{S} \right).
\end{equation}

The simulation used $\lambda=0.02$. Partial correlations smaller than $\tau_W=0.03$
in absolute magnitude were set to zero for topology-based outcomes. This
is ridge-like covariance regularisation followed by hard edge
thresholding, not graphical LASSO. Thresholded partial correlations were
used as an adjacency matrix W; they were not assumed automatically to
define a new positive-definite precision matrix. Because topology
outcomes depend strongly on regularisation and thresholding, applied
analyses should propagate network-estimation uncertainty through the
complete estimation--perturbation--scoring pipeline rather than relying
only on a small set of deterministic sensitivity scenarios.

\subsection{SymPerturb operator taxonomy, target anchors and estimands}\label{symperturb-operator-taxonomy-target-anchors-and-estimands}

SymPerturb distinguishes primitive perturbation operators from analytic
procedures. Virtual knockout, virtual knockdown, edge-level
communication blocking and node-centred communication blocking are
primitive operators. Virtual dosage perturbation evaluates an
intensity--response curve, combination perturbation constructs a joint
target and sequence optimisation searches an ordered decision objective.
All symptom variables should be oriented so that larger values represent
worse states. Let $\mathbf{c}_S$ denote clinically interpretable, symptom-specific
target anchors, which need not equal zero. A general location--scale map
is

\begin{equation*}\tag{M1}\label{eq:M1}
\mathbf{X}_{S}^{(\boldsymbol{\alpha})}
= \mathbf{c}_{S}
+ \mathbf{D}_{\mu}(\boldsymbol{\alpha})(\boldsymbol{\mu}_{S}-\mathbf{c}_{S})
+ \mathbf{D}_{\sigma}(\boldsymbol{\alpha})(\mathbf{X}_{S}-\boldsymbol{\mu}_{S}).
\end{equation*}

The corresponding target mean and covariance are $\mathbf{c}_S+\mathbf{D}_{\mu}(\boldsymbol{\alpha})(\boldsymbol{\mu}_S-\mathbf{c}_S)$ and $\mathbf{D}_{\sigma}(\boldsymbol{\alpha})\boldsymbol{\Sigma}_{SS}\mathbf{D}_{\sigma}(\boldsymbol{\alpha})$, respectively. Baseline recovery requires
$\mathbf{D}_{\mu}(0)=\mathbf{D}_{\sigma}(0)=\mathbf{I}$, whereas an exact knockout at the anchor requires
$\mathbf{D}_{\mu}(1)=\mathbf{D}_{\sigma}(1)=\mathbf{0}$. The present numerical implementation uses the linked
special case $\mathbf{c}_S=\mathbf{0}$ and $\mathbf{D}_{\mu}(\alpha)=\mathbf{D}_{\sigma}(\alpha)=\mathbf{D}_{\alpha}$. Sensitivity analyses
should compare location-only attenuation ($\mathbf{D}_{\sigma}=\mathbf{I}$), scale-only
attenuation ($\mathbf{D}_{\mu}=\mathbf{I}$) and independently parameterised location--scale
maps.

Three estimands should be distinguished. Direct target benefit evaluates
change in the perturbed targets S. Beneficial downstream spillover
evaluates improvement in K, whereas adverse spillover records worsening
in K. The current $G_j(\alpha)$ is a downstream-spillover estimand and
therefore does not represent total clinical benefit.

\subsection{Derivation of the location-scale state intervention}\label{derivation-of-the-location-scale-state-intervention}

Partition X, \ensuremath{\mu} and \ensuremath{\Sigma} into targets S and non-targets K. The Gaussian
conditional mean is

\begin{equation}\label{eq:14}
\mathbb{E}\left( \mathbf{X}_{K} \mid \mathbf{X}_{S} = \mathbf{x}_{S} \right) = \mathbf{\mu}_{K} + \mathbf{\Sigma}_{KS}\mathbf{\Sigma}_{SS}^{-1}\left( \mathbf{x}_{S} - \mathbf{\mu}_{S} \right).
\end{equation}

Define

\begin{equation}\label{eq:15}
\mathbf{B} = \mathbf{\Sigma}_{KS}\mathbf{\Sigma}_{SS}^{-1}
\end{equation}

and the residual

\begin{equation}\label{eq:16}
\mathbf{\varepsilon} = \mathbf{X}_{K} - \mathbf{\mu}_{K} - \mathbf{B}\left( \mathbf{X}_{S} - \mathbf{\mu}_{S} \right).
\end{equation}

Its cross-covariance with the targets is zero:

\begin{equation}\label{eq:17}
\operatorname{Cov}\left( \mathbf{\varepsilon},\mathbf{X}_{S} \right) = \mathbf{\Sigma}_{KS} - \mathbf{B}\mathbf{\Sigma}_{SS} = \mathbf{0}.
\end{equation}

Joint normality therefore implies independence, and the residual
covariance is the Schur complement \ensuremath{\Omega} defined in the Main text. The
intervention replaces the target component by

\begin{equation}\label{eq:18}
\mathbf{X}_{S}^{\left( \mathbf{\alpha} \right)} = \mathbf{D}_{\alpha}\mathbf{X}_{S},\qquad\mathbf{D}_{\alpha} = \operatorname{diag}\left( 1 - \alpha_{s} \right).
\end{equation}

This compact form is equivalent to the centred, zero-anchor linked
special case in the Main text because $\mathbf{D}_{\alpha}\mathbf{X}_S=\mathbf{D}_{\alpha}\boldsymbol{\mu}_S+\mathbf{D}_{\alpha}(\mathbf{X}_S-\boldsymbol{\mu}_S)$.
Substitution gives the non-target mean and covariance shown above. The
full post-intervention cross-covariance is

\begin{equation}\label{eq:19}
\operatorname{Cov}\left( \mathbf{X}_{K}^{\left( \mathbf{\alpha} \right)},\mathbf{X}_{S}^{\left( \mathbf{\alpha} \right)} \right) = \mathbf{B}\mathbf{D}_{\alpha}\mathbf{\Sigma}_{SS}\mathbf{D}_{\alpha}.
\end{equation}

At zero dose, $\mathbf{D}_{\alpha}=\mathbf{I}$ and all baseline moments are recovered. At unit
dose, $\mathbf{D}_{\alpha}=\mathbf{0}$, the target becomes a constant at the zero anchor and the
non-target covariance reduces to the Schur complement. These statements
describe the linked reference map; the general anchor and separate
location--scale maps are defined above.

\subsection{Bounded symptom scales}\label{bounded-symptom-scales}

The primary simulation treated symptom scores as approximately
continuous but restricted reported scores to $[0,4]$ using

\begin{equation}\label{eq:20}
h(y) = min\{ 4,\max(0,y)\}.
\end{equation}

For $Y\sim\mathcal{N}(m,s^2)$, let $a=(0-m)/s$ and $b=(4-m)/s$. Integration over the two
tails and the interior gives

\begin{equation}\label{eq:21}
\mathbb{E}\left[ h(Y) \right] = m\{\Phi(b) - \Phi(a)\} + s\{\phi(a) - \phi(b)\} + 4\{ 1 - \Phi(b)\}.
\end{equation}

When s=0, the expectation is h(m). This expression was used to calculate
population and fitted post-intervention means without Monte Carlo
sampling. It is a winsorised-normal expectation, not a truncated-normal
likelihood. The high-burden stress test increased means by 0.85, with a
maximum of 3.75, to expose boundary-induced dose curvature.

\subsection{Virtual knockout}\label{virtual-knockout}

For target j, vKO is the unit-dose state operator:

\begin{equation}\label{eq:22}
\mathcal{T}_{j}^{\mathrm{vKO}} = \mathcal{T}_{j,\alpha = 1}.
\end{equation}

In the numerical reference implementation, the target distribution is
degenerate at the zero anchor. State outcomes can be calculated
analytically because only the target variance is zero. If a
post-knockout GGM is required, the primary recommendation is to estimate
the induced network on $V\setminus\{j\}$. A topology-only knockout may instead set
all $w_{jk}=0$. A soft vKO can retain a pre-specified variance
$\varepsilon_j^2>0$, but $\varepsilon_j$ and the target anchor $c_j$ must be
reported and varied in sensitivity analysis.

\subsection{Virtual knockdown and dosage perturbation}\label{virtual-knockdown-and-dosage-perturbation}

Partial knockdown uses $0<\alpha<1$. For one target, the
unbounded Gaussian mean response is

\begin{equation}\label{eq:23}
\Delta\mathbf{\mu}_{-j}^{(\alpha)} = - \alpha\,\mathbf{\Sigma}_{-j,j}\sigma_{jj}^{-1}\mu_{j}.
\end{equation}

Consequently,

\begin{equation}\label{eq:24}
G_{j}(\alpha) = \alpha G_{j}(1)
\end{equation}

when the scale is unbounded, the covariance is fixed and the outcome is
linear in the non-target mean changes under the linked reference map.
Dose efficiency and responsiveness are then redundant with efficacy. The
implemented dose grid was \ensuremath{\alpha}\ensuremath{\in}\{0,0.10,0.25,0.50,0.75,1\}. Departure from
linearity was evaluated by comparing the bounded response with the
corresponding unbounded Gaussian response. Confirmatory analyses should
additionally compare location-only attenuation, scale-only attenuation
and the linked map.

\subsection{Edge-level and node-centred communication blocking}\label{edge-level-and-node-centred-communication-blocking}

For an undirected edge (u,v) with block fraction $q\in[0,1]$,

\begin{equation}\label{eq:25}
\mathbf{W}_{uv}^{(q)} = \mathbf{W} - q\, w_{uv}\left( \mathbf{e}_{u}\mathbf{e}_{v}^{\mathsf{T}} + \mathbf{e}_{v}\mathbf{e}_{u}^{\mathsf{T}} \right).
\end{equation}

Node-centred blocking attenuates all edges incident to j:

\begin{equation}\label{eq:26}
w_{ik}^{(j,q)}=
\begin{cases}
(1-q)w_{ik}, & i=j\ \text{or}\ k=j,\\
w_{ik}, & \text{otherwise}.
\end{cases}
\end{equation}

The reference block fraction was q=0.80. Because these operators act
directly on W, they evaluate graph topology unless the modified graph is
converted back to a valid precision matrix. Such coupling would require
preserving conditional-variance scales, enforcing positive definiteness
and re-inverting the precision matrix. The term ``communication'' is
therefore shorthand for topology attenuation and should not be read as a
biological transmission mechanism.

Finite-step propagation potential was defined as

\begin{equation}\label{eq:27}
Q_{T}\left( \mathbf{W} \right) = \mathbf{1}^{\mathsf{T}}\left\{ \sum_{t = 1}^{T}\left( \gamma\left| \mathbf{W} \right| \right)^{t} \right\}\mathbf{1}.
\end{equation}

The simulation used T=6 and \ensuremath{\gamma}=0.45. Absolute weights measure unsigned
route capacity irrespective of sign. Although finite T avoids a formal
infinite-series convergence requirement, the magnitude of $Q_T$ can still
be dominated by node degree, cycles and matrix scaling. A confirmatory
implementation should report the spectral radius of $\gamma|\mathbf{W}|$,
compare raw, row-normalised and spectral-normalised adjacency matrices,
and report signed and unsigned propagation functionals when inhibitory
relations are scientifically meaningful.

\subsection{Combination perturbation}\label{combination-perturbation}

For targets j and k, the joint operator applies unit dose to both
targets using the multivariate location-scale equations. To avoid
attributing direct removal of either target to one method but not the
other, all pair comparisons use the common outcome set $V\setminus\{j,k\}$. Let
$G_{jk}(1,1)$ be the mean standardised improvement on that set, and let
$G_j^{(-jk)}(1)$ and $G_k^{(-jk)}(1)$ be the corresponding single-target
benefits. Incremental pair value is

\begin{equation}\label{eq:28}
I_{jk} = G_{jk}(1,1) - max\{ G_{j}^{(-jk)}(1),G_{k}^{(-jk)}(1)\}.
\end{equation}

This contrast quantifies improvement beyond the better single target. It
is deliberately not called additive synergy, which would instead compare
the joint effect with the sum of single effects under an explicit causal
interaction model. The current exploratory implementation averages only
positive incremental values; confirmatory analyses should retain signed
pair values so that antagonistic or harmful combinations remain visible.
For each target, the reference partner set contained the five
highest-efficacy alternative targets.

\subsection{Sequence optimisation}\label{sequence-optimisation}

Let $\boldsymbol{\pi}=(\pi_1,\ldots,\pi_L)$ be an ordered sequence and
$S_t=\{\pi_1,\ldots,\pi_t\}$. A discounted sequence objective was

\begin{equation}\label{eq:29}
J(\pi) = \sum_{t = 1}^{L}\eta^{t - 1}\{ G\left( S_{t} \right) - G\left( S_{t - 1} \right)\} - \lambda_{c}\sum_{t = 1}^{L}c_{\pi_{t}},
\end{equation}

where $0<\eta\le1$ favours earlier benefit and $c_{\pi_t}$ is intervention
cost. Without discounting, order-dependent costs or constraints, or
adaptive model updating, the objective is order invariant for a fixed
final target set and reduces to subset selection. The sequence is
therefore induced by the decision objective, feasibility constraints and
costs---not by biological or temporal direction inferred from a
cross-sectional GGM.

\subsection{Seven utility outcomes and one robustness diagnostic}\label{seven-utility-outcomes-and-one-robustness-diagnostic}

For equal weights, standardised improvement was

\begin{equation}\label{eq:30}
\Delta_{i \leftarrow j}^{(\alpha)} = \frac{\mathbb{E}\left[ h\left( X_{i} \right) \right] - \mathbb{E}\left[ h\left( X_{i}^{(j,\alpha)} \right) \right]}{\sqrt{{\widehat{\sigma}}_{ii,\lambda}}}.
\end{equation}

The target itself was excluded from every downstream system-benefit,
breadth and single-versus-pair comparison. The current $G_j(\alpha)$ is
therefore a downstream-spillover estimand rather than total clinical
benefit. Applied analyses should report direct target benefit,
beneficial downstream spillover and adverse downstream spillover
separately. Reference thresholds were \ensuremath{\tau}=0.10 standard deviations for
breadth and $\tau_m=0.20$ for a module-average improvement. Dose efficiency
used $\alpha_k\in\{0.25,0.50,0.75\}$, responsiveness used $\varepsilon=0.10$, and
communication blocking used the finite-step topology definition above.

The seven utility outcomes and the separate robustness diagnostic were
calculated as

\begin{equation}\label{eq:31}
R_{j}^{\mathrm{eff}} = G_{j}(1),
\end{equation}

\begin{equation}\label{eq:32}
R_{j}^{\mathrm{dose}} = \frac{1}{K}\sum_{k = 1}^{K}\frac{G_{j}\left( \alpha_{k} \right)}{\alpha_{k}},
\end{equation}

\begin{equation}\label{eq:33}
R_{j}^{\mathrm{breadth}} = \frac{1}{p - 1}\sum_{i \neq j}\mathbb{I}\left( \Delta_{i \leftarrow j}^{(1)} \geq \tau \right),
\end{equation}

\begin{equation}\label{eq:34}
R_{j}^{\mathrm{cross}} = \frac{1}{q - 1}\sum_{m \neq c(j)}\mathbb{I}\left( {\bar{\Delta}}_{m \leftarrow j}^{(1)} \geq \tau_{m} \right),
\end{equation}

\begin{equation}\label{eq:35}
R_{j}^{\mathrm{rob}} = 1 - \frac{{\mathrm{SD}}_{\theta}\{{\mathrm{rank}}_{j}^{(\theta)}\}}{\max_{\ell}{\mathrm{SD}}_{\theta}\{{\mathrm{rank}}_{\ell}^{(\theta)}\}},
\end{equation}

\begin{equation}\label{eq:36}
R_{j}^{\mathrm{comm}} = \frac{Q_{T}\left( \mathbf{W} \right) - Q_{T}\left( \mathbf{W}_{j}^{\mathrm{block}} \right)}{Q_{T}\left( \mathbf{W} \right)},
\end{equation}

\begin{equation}\label{eq:37}
R_{j}^{\mathrm{comb}} = \frac{1}{\left| \mathcal{T}_{j} \right|}\sum_{k \in \mathcal{T}_{j}}\max\left[ 0,\, G_{jk}(1,1) - max\{ G_{j}^{(-jk)}(1),G_{k}^{(-jk)}(1)\} \right],
\end{equation}

\begin{equation}\label{eq:38}
R_{j}^{\mathrm{resp}} = \frac{G_{j}(\epsilon) - G_{j}(0)}{\epsilon}.
\end{equation}

Robustness was calculated from a provisional composite of the other
seven normalised outcomes. The 13 scenarios were: reference settings;
ridge values 0 and 0.05; breadth thresholds 0.05 and 0.15; module
thresholds 0.10 and 0.30; propagation decay 0.35 and 0.55; block
fractions 0.60 and 1.00; and partner-set sizes 3 and 8. The standard
deviation of each target rank across these scenarios was converted to
the robustness diagnostic in Table 2. Because this quantity was weakly
recoverable and reflects uncertainty rather than utility, it is excluded
from the revised confirmatory VPPS. The numerical results in the current
draft were generated before this revision and require reanalysis.

\subsection{Simulation design}\label{simulation-design}

The generating network had 22 named symptom nodes and four modules:
Somatic, Affective, Sleep--fatigue and Cognitive--behavioural.
Within-module edges were sampled more densely and strongly than
cross-module edges. Additional bridge edges connected Fatigue with
Functional limitation and Concentration, Worry with Insomnia, Pain with
Insomnia, Sadness with Social withdrawal, Anxiety with Palpitations, Low
energy with Anhedonia and Rumination with Insomnia. A minority of edges
were negative. The symmetric partial-correlation matrix was scaled to
spectral radius at most 0.68, and

\begin{equation}\label{eq:39}
\mathbf{\Theta} = \mathbf{I} - \mathbf{W},\qquad\mathbf{\Sigma}_{corr} = \operatorname{cor}\left( \mathbf{\Theta}^{-1} \right).
\end{equation}

Node standard deviations ranged from 0.60 to 0.84 and means ranged from
1.30 to 2.80. The covariance was $\operatorname{diag}(\mathbf{s})\boldsymbol{\Sigma}_{\mathrm{corr}}\operatorname{diag}(\mathbf{s})$. Each dataset was
sampled from the resulting multivariate normal distribution and clipped
to $[0,4]$. The full parameter values and a generated n=1,000 dataset
accompany the code.

Population outcomes were computed from the known generating parameters.
At each sample size, 200 independent datasets were generated from
distinct child seeds of the master seed 20260727. Sample means and
covariances were then passed through the same fitting and scoring
functions. The primary recovery statistic was Spearman correlation
between estimated and population target ranks. We also calculated the
proportion of the population top five recovered and root mean squared
error of the original exploratory VPPS on its 0--100 scale. This design
evaluates finite-sample recovery under correct model specification; it
does not assess robustness to misspecification.

The analytical-versus-Monte Carlo check used 100,000 post-vKO draws for
each of 22 targets. The vKO validity check used 100 independently
generated post-intervention datasets of size 500 and compared exact
full-network vKO, exact vKO with the target removed, and soft vKO
retaining 5\% of the target standard deviation. Parameter-sensitivity
correlations compared each scenario with the reference population rank.

\subsection{Required confirmatory validation extensions}\label{required-confirmatory-validation-extensions}

The numerical results in this draft provide internal verification under
a single model-compatible GGM family. A general methods claim requires a
factorial stress-test suite that varies data type (continuous, ordinal
and zero-inflated), latent confounding, measurement error, missingness,
network density, edge-sign balance, module strength, p/n ratio,
estimator and regularisation or thresholding strategy. Performance
should be summarised across factors rather than within one favourable
generating mechanism.

SymPerturb should be benchmarked against absolute strength, expected
influence, bridge strength or bridge expected influence, predictability,
control-based metrics and a burden-only rule. The comparison should
pre-specify losses such as population-rank recovery, top-k recovery,
expected decision regret, stability and calibration of predicted
downstream change.

A complete-pipeline bootstrap should repeat network estimation,
threshold selection, perturbation, outcome calculation, candidate-set
normalisation and ranking. An empirical benchmark should then test
whether longitudinal or experimentally induced target changes precede
the predicted downstream changes and whether the intervention
demonstrably engages the intended target.

\subsection{Statistical analysis}\label{statistical-analysis}

No null-hypothesis significance tests were used. The independent unit
for finite-sample verification was an independently generated dataset.
Medians and 2.5th to 97.5th percentile Monte Carlo intervals summarised
the 200 replicate distributions. These intervals quantify
simulation-to-simulation variability under the specified generating
model; they are not confidence intervals for patients, diseases or
treatment effects. Monte Carlo draws used to verify an analytical
expectation were not treated as additional independent datasets. For
applied analyses, nonparametric or parametric bootstrap resampling
should repeat network estimation, thresholding, perturbation, outcome
calculation, normalisation and ranking in every replicate, yielding
uncertainty intervals, rank distributions and top-k selection
probabilities.

\subsection{Software and reproducibility}\label{software-and-reproducibility}

The supplied R implementation defined the original symptom labels,
module structure and conceptual perturbation workflow. The revised
validation was executed in Python 3.12 using NumPy, pandas, SciPy and
Matplotlib because an R runtime was not available in the execution
environment. The accompanying script contains data generation, the
original eight outcomes, VPPS calculation, validation and figure export.
Before submission, the software must be updated to implement
symptom-specific anchors, separate location and scale maps, signed
combination values, a seven-utility-dimension VPPS, separate robustness
diagnostics and complete-pipeline bootstrap uncertainty. An
independently reviewed implementation in the intended analysis
environment should reproduce all source-data tables.

Generative artificial intelligence was used to assist manuscript
drafting, mathematical cross-checking and code review. All equations,
generated data, numerical outputs and interpretations require author
verification, and accountability remains with the named authors.

\section*{Data availability}\label{data-availability}

All data used in this draft are synthetic. The generated n=1,000 example
dataset, population parameters, source data underlying the validation
results and target-level scores accompany the reproducibility bundle.
Before submission, these files should be deposited in a public research
repository with a persistent identifier. No patient data were used.

\section*{Code availability}\label{code-availability}

The core computational code underlying the original virtual-perturbation
implementation has been published previously in a related methodological
article. The present study extends that implementation by introducing
the SymPerturb framework, including general target anchors, separate
location--scale mappings, seven utility outcomes, robustness as a
separate uncertainty diagnostic and full-pipeline uncertainty
assessment. The revised code used for the analyses reported here,
together with machine-readable outputs and source data, will be made
available in a public, version-controlled repository upon publication.
The previously published R code will be retained as methodological
provenance, whereas the updated SymPerturb definitions described in this
Article will govern the validated implementation.

\section*{Acknowledgements}\label{acknowledgements}

This work was supported by the National Natural Science Foundation of
China (Grant No. 72574043), the Shanghai Pujiang Program (Grant No.
24PJC014), and the China University Industry--Research Innovation
Fund---Digital Intelligence Innovation and Talent Program (Grant No.
2024LC007), all awarded to Z.Z. The funders had no role in the study
design, data collection, data analysis, interpretation of the results,
decision to publish, or preparation of the manuscript.

\section*{Author contributions}\label{author-contributions}

Z.Z. conceived and developed the SymPerturb framework, designed the
methodological study, supplied the original R implementation and
evaluation specifications, and drafted the manuscript. J.Y., T.H. and
Z.Y. provided methodological and conceptual feedback and critically
reviewed and revised the manuscript. All authors reviewed and approved
the final version of the manuscript.

\section*{Competing interests}\label{competing-interests}

The authors declare no competing interests.

\nocite{*}
\bibliographystyle{unsrt}
\bibliography{references}

\end{document}